\documentclass[11pt]{article}
\usepackage[blocks]{authblk}
\usepackage{amsmath,amssymb,graphicx,url}
\usepackage{color}

\date{\today}

\title{A note on a better conditioned Domain Wall Operator}

\author{H.\ Neff, Luzernerstrasse 43, 6330 Cham, Switzerland \footnote{Email: hartmutneff@aol.com}}

\begin{document}

\maketitle

\begin{abstract}
  This is a brief note on 'A better conditioned Domain Wall
  Operator', which provides a more detailed explanation of the domain wall-to-overlap transformation with the inclusion of the $\alpha$ parameter.
\cite{Chen:2026, Neff:2015}.
\end{abstract}


\section{The better conditioned Domain Wall operator}

The better conditioned Domain Wall operator introduces a parameter
$\alpha$, and takes on the following form, presented for $L_s = 4$ to keep notation simple:
\begin{eqnarray}\label{eq:mobius1}
D_{\alpha}(m) =
 &\hskip-0.4cm\left[\hskip-0.2cm\begin{array}{ccccc}
 D_{1+} (P_- + \alpha P_+ )  & \alpha  D_{1-} P_- &   0  &    -mD_{1-} P_+ \\
 \alpha D_{2-} P_+  &   \alpha D_{2+}  &   \alpha D_{2-} P_-  &  0  \\
 0  &   \alpha D_{3-} P_+  &  \alpha  D_{3+}  &   \alpha D_{3-} P_-  \\
 -mD_{4-} P_-  &   0  & \alpha D_{4-} P_+ &  D_{4+} (P_+ + \alpha P_- ) \\
\end{array} \hskip-0.2cm \right] &
\end{eqnarray}
with
\begin{eqnarray}
&D_{i+} = b_i D_w +1 , \;\;\; D_{i-} = c_i D_w -1 \label{eq:coeff} ,& \\
&P_+ = \frac{1}{2} ( 1 + \gamma_5) , \;\;\; P_- = \frac{1}{2} ( 1 - \gamma_5) .&
\end{eqnarray}
$D_w$ denotes the Wilson Dirac matrix
\begin{eqnarray}\label{eq:wilson}
D_w (M_5)\hskip-0.05cm = \hskip-0.05cm (4+M_5) \delta_{x,y} -\hskip-0.1cm \frac{1}{2}  \bigl[  (1 \hskip-0.05cm   - \hskip-0.05cm   \gamma_\mu) 
 U_\mu(x) \delta_{x+\mu,y}  
\hskip-0.05cm  +  \hskip-0.05cm  (1  \hskip-0.05cm  +  \hskip-0.05cm   \gamma_\mu) U_\mu^\dagger(y) \delta_{x,y+\mu}  \bigr] .
\end{eqnarray}

An equivalent form of the 5D operator can be obtained by multiplying
it from the right by a matrix of chiral projectors. This form makes
the changes introduced by $\alpha$ more obvious and trivial: it amounts to
merely scaling some of the matrix columns. Multiplying equation
eq.(\ref{eq:mobius1}) from the right with P (see eq.(\ref{eqP}))
leads to

\begin{eqnarray} 
&D_{\alpha} P = D_{\alpha} 
\left[\begin{array}{ccccc}
 P_- &   P_+  &   0 &   0  \\
 0 &   P_-  &   P_+ &   0 \\
 0 &   0  &   P_-  &   P_+  \\
  P_+ &   0  &    0 &   P_- \\
  \end{array} \right] & \\
&=\gamma_5  \left[\begin{array}{ccccc}
      Q_{1-} c_-  & \alpha  Q_{1+} & 0 & 0  \\
      0 &  \alpha Q_{2-} & \alpha  Q_{2+}& 0  \\
      0 &  0& \alpha Q_{3-} &   \alpha  Q_{3+} \\
      Q_{4+} c_+ & 0  & 0& \alpha Q_{4-}   \\
  \end{array} \right] &
\end{eqnarray}

The matrix entries are defined as,
\begin{eqnarray}
 & Q_{i+} = \gamma_5 D_w (b_i P_+ + c_i P_-) +1,\;\;\; Q_{i-} = \gamma_5 D_w (b_i P_- + c_i P_+) -1,& \nonumber\\
& c_+ = P_+ - m P_- ,\;\;\; 
 c_- = P_- - m P_+ . &
\end{eqnarray}

The mass only appears in $c_ + $ and $c_ - $, and therefore only in
the first column. The scaling factor, $\alpha$, is applied to all the other
columns, the ones containing no mass term. This suggests that this
approach might be especially helpful for small quark masses.

Let's put this operator in the standard Domain Wall to Overlap
transformation to understand what $\alpha$ does to the 4D propagator.

\section{Domain Wall to Overlap transformation}
\label{sec:appa}

  The Domain Wall to Overlap transformation reads,
  \begin{eqnarray}\label{eq:4d-5d}
    L   D_{DW}(m) R(m)  =  F  D^5_{OV}(m).
\end{eqnarray}
The transformation matrices take the form (for $L_s = 4$ sites in the 5th dimension),
\begin{eqnarray}
F  = L   D_{DW}(1)  R(1) ,
\end{eqnarray}
and
\begin{eqnarray}
L \hskip-0.05cm = \hskip-0.05cm  L_1 L_2 \hskip-0.05cm  = \hskip-0.05cm 
\left[\begin{array}{cccc}
1 & S_1   & S_1 S_2    & S_1 S_2 S_3   \\
0 & 1        & S_2           & S_2 S_3   \\
0 & 0        & 1                & S_3   \\
0 & 0        & 0                & 1 
\end{array} \right] \hskip-0.1cm
\left[\begin{array}{cccc}
Q_{1-}^{-1}  & 0  & 0 & 0  \\
0 & Q_{2-}^{-1}    & 0 & 0  \\
0 & 0  & Q_{3-}^{-1}   & 0  \\
0 & 0  & 0 & Q_{4-}^{-1}   \\
\end{array} \right]\gamma_5  , \nonumber
\end{eqnarray}\begin{eqnarray} \label{eqP}\nonumber
R(m) = P R_1(m) =
\left[\begin{array}{cccc}
 P_- &   P_+  &   0 &   0  \\
 0 &   P_-  &   P_+ &   0  \\
 0 &   0  &   P_- &   P_+  \\
 P_+ &   0  &   0 &   P_- \\
\end{array} \right] 
\left[\begin{array}{cccc}
  -1 &   0  &   0 &   0  \\
  - \frac{1}{\alpha} S_2 S_3 S_4 \, c_+ &   1  &   0 &   0  \\
  - \frac{1}{\alpha} S_3 S_4 \, c_+ &   0  &   1 &   0  \\
  - \frac{1}{\alpha} S_4 \, c_+ &   0  &   0 &   1 \\
\end{array} \right],
\end{eqnarray}
\begin{eqnarray}
&& D^5_{OV}(m)= \left[\begin{array}{cccc}
D^4_{OV}(m) & 0  & 0 & 0  \\
0 & 1  & 0 & 0  \\
0 & 0  & 1 & 0  \\
0 & 0  & 0 & 1 \\
\end{array} \right].
\end{eqnarray}

$T_i^{-1}$ is called the transfer matrix.

\begin{equation}
S_i = T_i^{-1} = - Q_{i-}^{-1} Q_{i+} ,\nonumber
\end{equation}
The matrix multiplications will be performed in the following order,
\begin{eqnarray}
L_1 L_2 D_{DW}(m) P R_1(m) = L_1 L_2 M_1 R_1(m) =  L_1 M_2 R_1(m) = L_1 M_3 = M_4 .
\end{eqnarray}
Step 1:
  \begin{eqnarray}
M_1  = D_{DW}(m) P 
   =\gamma_5  \left[\begin{array}{cccc}
      Q_{1-} c_-  &   \alpha Q_{1+} & 0 & 0  \\
      0 &   \alpha Q_{2-} &  \alpha  Q_{2+} & 0  \\
      0 & 0  & \alpha Q_{3-} &   \alpha Q_{3+}  \\
      Q_{4+} c_+ & 0  & 0 &  \alpha Q_{4-}   \\
    \end{array} \right],
  \end{eqnarray}
  with
  \begin{eqnarray}
   Q_{i-} & =&  \gamma_5 (D_{i+} P_- + D_{i-} P_+) \nonumber\\
    &=& \gamma_5( D_w (b_i P_- + c_i P_+) +P_- - P_+) \nonumber\\
    &=& \gamma_5 D_w(b_i P_- + c_i P_+) - 1,\\
    Q_{i+} &=& \gamma_5(D_{i+} P_+ + D_{i-} P_-)\nonumber\\
    & =& \gamma_5  ( D_w (b_i P_+ + c_i P_-) +P_+ - P_-)\nonumber\\
    &=& \gamma_5 D_w(b_i P_+ + c_i P_-) + 1 .
  \end{eqnarray}
Step 2:
  \begin{eqnarray}
  M_2 =  L_2 M_1 =
    \left[\begin{array}{cccc}
      c_-  & - \alpha S_1 & 0 & 0  \\
      0 &  \alpha   & - \alpha S_2 & 0  \\
      0 & 0  &  \alpha  & - \alpha S_3 \\
      -S_4 c_+ & 0  & 0 &  \alpha    \\
    \end{array} \right].
  \end{eqnarray}
Step 3:
  \begin{eqnarray}
  M_3 =  M_2 R_1(m)=
    \left[\begin{array}{cccc}
      -c_- + S_1 S_2 S_3 S_4 c_+  & - \alpha S_1 & 0 & 0  \\
      0 & \alpha  & - \alpha S_2 & 0  \\
      0 & 0  & \alpha  & - \alpha S_3 \\
      0 & 0  & 0 &  \alpha   \\
    \end{array} \right].
  \end{eqnarray}
Step 4:
  \begin{eqnarray}
  L   D_{DW}(m)  R(m)= M_4 = L_1 M_3 =
    \left[\begin{array}{cccc}
      -c_- + S_1 S_2 S_3 S_4 c_+  & 0 & 0 & 0  \\
      0 &  1  & 0 & 0  \\
      0 & 0  & 1  & 0\\
      0 & 0  & 0 & 1   \\
    \end{array} \right].
  \end{eqnarray}
This leads to,
  \begin{eqnarray} 
 F = L D_{DW}(1) R(1) =
    \left[\begin{array}{cccc}
      (1 + S_1 S_2 S_3 S_4 ) \gamma_5 & 0 & 0 & 0  \\
      0 &  1   & 0 & 0  \\
      0 & 0  & 1   & 0\\
      0 & 0  & 0 & 1    \\
    \end{array} \right].
  \end{eqnarray}
To make notation simpler, we define $S=S_1 S_2 S_3 S_4 $. The 5D Overlap Operator takes the form,
  \begin{eqnarray}\label{startprop}
 D^5_{OV}(m) = F^{-1} M_4  = 
    \left[\begin{array}{cccc}
      \gamma_5 (1 + S )^{-1} (-c_- + S c_+)  & 0 & 0 & 0  \\
      0 &  1  & 0 & 0  \\
      0 & 0  & 1   & 0\\
      0 & 0  & 0 & 1    \\
    \end{array} \right].
  \end{eqnarray}
It follows for the $(11)$ element,
\begin{eqnarray}
  D^5_{OV}(m)_{11} &=& \frac{1}{2}\gamma_5 (1 + S)^{-1}   \left( m + m \gamma_5 -1 + \gamma_5 + S (1+\gamma_5 -m +m\gamma_5)\right)\nonumber\\
&=&  \frac{1}{2}  \gamma_5(1 + S )^{-1}   \left( (1+m) (S+1) \gamma_5 +  (1-m) (S - 1)\right)\nonumber\\
&=&  \frac{1}{2}  \left( (1+m)   +  (1-m) \gamma_5\frac{ (S - 1)}{(S+1)}\right).
\end{eqnarray}
Hence eq.(\ref{startprop}) takes the form, 
  \begin{eqnarray}
 D^5_{OV}(m) =
    \left[\begin{array}{cccc}
      \frac{1}{2}  \left( (1+m)  +  (1-m) \gamma_5\frac{ (S - 1)}{(S+1)}\right) & 0 & 0 & 0  \\
      0 &  1  & 0 & 0  \\
      0 & 0  & 1  & 0\\
      0 & 0  & 0 & 1  \\
    \end{array} \right].
  \end{eqnarray}
  The matrix that acts as the variable for the polar decomposition can be found by setting,
    \begin{eqnarray}\label{polar}
    \frac{ (S-1)}{(S+1)} = \frac{(1-1/S)}{(1+1/S)}=
    \frac{\Pi_{i=1}^4(1+a_iX_i) -
      \Pi_{i=1}^4(1-a_iX_i)}{\Pi_{i=1}^4(1+a_iX_i) + \Pi_{i=1}^4(1-a_iX_i)},
    \end{eqnarray}
     and therefore
      \begin{eqnarray}\label{polar1}
         \frac{1}{S} = \frac{1}{S_1 S_2 S_3 S_4} =
        \frac{(1-a_1X_1(1-a_2X_2)(1-a_3X_3)(1-a_4X_4)}{(1+a_1X_1)(1+a_2X_2)(1+a_3X_3)(1+a_4X_4)}.
      \end{eqnarray}
      For each $i$, we determine $X_i$,
         \begin{eqnarray}
        & S_i^{-1} =
           (1-a_iX_i)(1+a_iX_i)^{-1}&\nonumber\\
           &Q_{i-}^{-1} Q_{i+} = (a_iX_i+1)(a_iX_i-1)^{-1}&\nonumber\\
         &    Q_{i+} (a_iX_i-1) = Q_{i-}(a_iX_i+1)&\nonumber\\
           &  a(Q_{i+} -Q_{i-}) X_i =  Q_{i+} +Q_{i-} & \nonumber\\
           & a_i \gamma_5 ( (b_i-c_i)  D_w + 2  )\gamma_5 X_i = (b_i+c_i) \gamma_5 D_w .&
     \end{eqnarray}
         This results in,
         \begin{eqnarray}\label{argument}
       a_i X_i = (b_i+c_i) \gamma_5 D_w \frac{1}{2+(b_i-c_i)D}.
         \end{eqnarray}
We can therefore write,
         \begin{eqnarray}
 D^5_{OV}(m) =
    \left[\begin{array}{cccc}
       D^4_{OV}(m) & 0 & 0 & 0  \\
      0 &  1  & 0 & 0  \\
      0 & 0  & 1  & 0\\
      0 & 0  & 0 & 1  \\
    \end{array} \right].
  \end{eqnarray}
     
\section{Computation of the 4D propagator}
\label{sec:appb}

It follows directly from,
  \begin{eqnarray}
D^5_{OV}(m) = 
    \left[\begin{array}{cccc}
      D_{OV}^4  & 0 & 0 & 0  \\
      0 &  1  & 0 & 0  \\
      0 & 0  & 1   & 0\\
      0 & 0  & 0 & 1    \\
    \end{array} \right]
    \left(\begin{array}{c} x_1 \\ x_2\\x_3\\x_4 \end{array} \right)
    =     \left(\begin{array}{c} b_1 \\ b_2\\b_3\\b_4 \end{array} \right),
  \end{eqnarray}
  or
    \begin{eqnarray}
  D_{OV}^4  x_1 = b_1 ,
    \end{eqnarray}
    that the 4D propagator is equal to $x_1$. We use eq.(\ref{eq:4d-5d}) and find
      \begin{eqnarray}
         F^{-1} L D_{DW}(m) R(m) \vec{x} = \vec{b} ,
      \end{eqnarray}
or
      \begin{eqnarray}
     D^5_{OV}(m) \vec{x} =     R_1^{-1}(1) P^{-1} D_{DW}^{-1}(1) D_{DW}(m) P R_1(m) \vec{x} = \vec{b} .
      \end{eqnarray}
      Together with $R_1^{-1} = R_1$, it follows from,
       \begin{eqnarray}\label{finalEq}
   R_1(1) D^5_{OV} R_1^{-1}(m) \vec{y} =
     \left[\begin{array}{cccc}
         D^4_{OV} & 0 & 0 & 0  \\ 
       \frac{1}{\alpha} S_2 S_3S_4 (\gamma_5 D^4_{OV} -c_+) & 1  & 0 & 0  \\
       \frac{1}{\alpha} S_3 S_4 (\gamma_5 D^4_{OV} -c_+)  & 0  & 1  & 0\\
       \frac{1}{\alpha} S_4 (\gamma_5 D^4_{OV} -c_+)  & 0  & 0 & 1   \\
    \end{array} \right] \vec{y} = \vec{b},
  \end{eqnarray}
       that $y_1 = x_1$, i.e.\ that the 4D propagator is unaffected
       by the transformation with $R_1$ or by $\alpha$.  Hence we can use

       \begin{eqnarray}
      R_1(1) D^5_{OV} R_1^{-1}(m) \vec{y} =  P^{-1} D_{DW}^{-1}(1) D_{DW}(m) P  \vec{y} = \vec{b} ,
    \end{eqnarray}
or
    \begin{eqnarray}\label{tosolve}
      D_{DW}(m) P  \vec{y} =D_{DW}(1) P \vec{b} ,
    \end{eqnarray}
    to determine the 4D propagator $y_1$.

    It should be obvious from eq.(\ref{finalEq}) that $\alpha$ does
    not affect the 4D propagator.  Another observation is that one
    could choose different values of $\alpha$s for each column of the Domain
    Wall matrix, such as $1, \alpha_1, \alpha_2, \alpha_3, \ldots, ,
    \alpha_{L_s}$. It turned out, in the test that were
    performed, to be numerically optimal to choose all $\alpha$s
    equal.

    One could look at this in a perhaps much simpler way, by dividing
    the Domain Wall Matrix with $\alpha$, which would replace
    $c_+$ and $c_-$ with $\frac{c_+}{\alpha}$ and
    $\frac{c_-}{\alpha}$. This again emphasises, that $\alpha$ almost
    directly scales the quark masses and hence that this method might be
    especially useful for when they are small.

    It is important to note that, if even-odd preconditioning were to
    be applied, it would have to be done from the left in order not to
    cancel out the effect of $\alpha$.

    \bibliography{note}       
\end{document}